\documentstyle[amssymb,12pt,thmsa,sw20rui]{article}

\input{tcilatex}

\begin{document}

\author{Hongya Liu$\thanks{%
Email: hyliu@dlut.edu.cn}\smallskip $ \\
Department of Physics, Dalian University of Technology, \\
Dalian, 116024, P.R. China\smallskip \and Paul S. Wesson$\thanks{%
Email: wesson@astro.uwaterloo.ca}$ \\
Department of Physics, University of Waterloo, \\
Waterloo, Ontario N2L 3G1, Canada}
\title{Radiating sources in higher-dimensional gravity}
\maketitle

\begin{abstract}
We study a time-dependent 5D metric which contains a static 4D sub-metric
whose 3D part is spherically symmetric. An expansion in the metric
coefficient allow us to obtain close-to Schwarzschild approximation to a
class of spherically-symmetric solutions. Using Campbell's embedding theorem
and the induced-matter formalism we obtain two 4D solutions. One describes a
source with the stiff equation of state believed to be applicable to dense
astrophysical objects, and the other describes a spherical source with a
radial heat flow.

PACS number(s): 04.20.Jb, 11.10 Kk
\end{abstract}

\section{INTRODUCTION}

Campbell showed the theorem that any solution of the Einstein equations in $%
N $ dimensions can be locally embedded in a Ricci-flat manifold of $(N+1)$
dimensions whose field equations in terms of the Ricci tensor are $R_{AB}=0$
($A,B=0,1,...,N$).$^1$ Tavakol and coworkers have recently noted the
relevance of this to the embedding of lower-dimensional (possibly
quantizeable) gravity in 4D Einstein gravity,$^2$ and the technique is
clearly applicable to the recovery of solutions of 4D general relativity
from 10D superstrings, 11D supergravity and M-theory.$^{3-6}$ A major
application of Campbell's theorem is to induced-matter theory, wherein
solutions of the 4D Einstein equations with matter are recovered from the 5D
Kaluza-Klein equations in apparent vacuum.$^7$ This approach has been
applied to cosmology,$^8$ clusters of galaxies$^9$ and the solar system,$%
^{10,11}$ where there is agreement with observational data. It is valuable
in application to general relativity, since solutions of the Kaluza-Klein
equations can yield new solutions of the Einstein equations. With regard to
the latter, spherically-symmetric sources such as stars can be modelled in
the simplest way by the interior and exterior Schwarzschild solutions. But
to include the radiation outside a star, more complicated solutions are
required.$^{12}$ These include the Vaidya metric which uses a retarded time
coordinate to describe a radiating atmosphere,$^{13}$ the metrics of Herrera
and coworkers wherein spheres of matter are matched to exterior spacetimes,$%
^{14-16}$ and the metrics of Glass and Krisch which extend the Vaidya
solution to include both a radiation field and a string fluid.$^{17}$
However, despite extensive work on star-like solutions of the 4D equations
and the existence of Campbell's theorem which shows that such can be
embedded in the $N$D equations, not much work has been done on
spherically-symmetric solutions of the 5D equations. Notable exceptions are
solutions which have an isothermal equation of state in 4D$^9$ and ones
which are flat in 5D but curved in 4D.$^{18}$ We will therefore present an
analysis of a class of 5D metrics, and illustrate their relevance by
isolating two 4D solutions. One describes a source with the stiff equation
of state believed to be applicable to dense astrophysical objects, and the
other describes a spherical source with radiation.

\section{5D METRICS WITH 4D SPHERICAL SOURCES}

In this section we let upper-case Latin indices run 0-4 and lower-case Greek
indices run 0-3. We absorb the speed of light and the gravitational constant
through the choices of units $c=1$, $8\pi G=1$. The coordinates are $%
x^A=t,r,\theta ,\phi ,y$ with $d\Omega ^2\equiv d\theta ^2+\sin ^2\theta
d\phi ^2$ with (3D) spherical symmetry.

Consider the 5D line element with interval 
\begin{eqnarray}
dS^2 &=&g_{AB}dx^Adx^B=\Phi ^2ds^2-\Phi ^{-4}dy^2\;, \\
ds^2 &=&g_{\alpha \beta }dx^\alpha dx^\beta \;\;,
\end{eqnarray}
where $g_{\alpha \beta }=g_{\alpha \beta }(x^\mu )$ and $\Phi =\Phi (x^\mu )$
.. This metric is broad in scope and has been used in other contexts.$^{19}$
It also has the advantage that the 5D field equations $R_{AB}=0$ break down
neatly into a set involving the 4D Ricci tensor and a conservation-type
equation for the scalar field $(g_{44}=-\Phi ^{-4})$:

\begin{eqnarray}
R_{\alpha \beta } &=&6\Phi ^{-2}\Phi _\alpha \Phi _\beta \quad , \\
\Phi \Phi _{;\alpha }^\alpha -\Phi ^\alpha \Phi _\alpha &=&0\;\;.
\end{eqnarray}
Here $\Phi ^\alpha =g^{\alpha \beta }\Phi _\beta ,$ $\Phi _{;\alpha }^\alpha
=g^{\alpha \beta }\Phi _{\alpha ;\beta }$ and $R_{\alpha \beta }$ in (3) is
made of $g_{\alpha \beta }$. Equations (3) and (4) admit many different
types of solution, but here we choose a time-dependent scalar field and a
static 4D geometry:

\begin{eqnarray}
\Phi &=&\Phi (t)\quad , \\
ds^2 &=&B(r)dt^2-A(r)dr^2-r^2d\Omega ^2\quad .
\end{eqnarray}
These require that the LHS of (3) should be independent of $t$ and the RHS
of (3) should be independent of $r$. Therefore both sides should equal to a
constant, say, $6\lambda ^2$. So we must have 
\begin{equation}
\Phi =e^{\lambda t}\;\;.
\end{equation}
Substituting this into equation (4), we find that (4) is satisfied. Using
(6) and (7) in (3) gives the non-vanishing components

\begin{eqnarray}
R_{00} &=&\frac{B^{\prime \prime }}{2A}-\frac{B^{\prime }}{4A}\left( \frac{%
A^{\prime }}A+\frac{B^{\prime }}B\right) +\frac 1r\frac{B^{\prime }}%
A=6\lambda ^2\quad , \\
R_{11} &=&-\frac{B^{\prime \prime }}{2B}+\frac{B^{\prime }}{4B}\left( \frac{%
A^{\prime }}A+\frac{B^{\prime }}B\right) +\frac 1r\frac{A^{\prime }}%
A=0\;\quad , \\
R_{22} &=&\sin ^{-2}\theta R_{33}=1-\frac 1A+\frac r{2A}\left( \frac{%
A^{\prime }}A-\frac{B^{\prime }}B\right) =0\;,
\end{eqnarray}
with $A^{\prime }\equiv dA/dr$. These are three equations in two unknowns,
and determine a class of solutions which is time-dependent in 5D via (7) but
static in 4D and spherically symmetric in 3D.

Let us manipulate (8)-(10). The sum of $AB^{-1}R_{00}$ of (8) and $R_{11}$
of (9) gives

\begin{equation}
\frac{A^{\prime }}{A}+\frac{B^{\prime }}{B}=6\lambda ^{2}r\frac{A}{B}\;\;.
\end{equation}
And we can rewrite (10) as

\begin{equation}
\frac{A^{\prime }}A-\frac{B^{\prime }}B=-\frac 2r\left( A-1\right) \;\;.
\end{equation}
Using (11) and (12), we verify that (8) and (9) are satisfied. Therefore, we
need to solve (11) and (12) for $A$ and $B$. Now (12) can be written as $%
(d/dr)\ln (A/B)=-2(A-1)/r$. Integrating this we get

\begin{equation}
B=A\exp \left[ 2\int_{r_0}^r\frac{A-1}rdr\right] \quad .
\end{equation}
Substituting this into (11), we obtain

\begin{equation}
\frac{A^{\prime }}A+\frac{A-1}r=3\lambda ^2r\exp \left[ -2\int_{r_0}^r\frac{%
A-1}rdr\right] \quad .
\end{equation}
Without loss of generality, let us introduce a mass function$^{20,21}$ via

\begin{equation}
A=\left( 1-\frac{2\mu (r)}r\right) ^{-1}\quad .
\end{equation}
Then (14) becomes 
\begin{equation}
\mu ^{\prime }=\frac 32\lambda ^2r\left( r-2\mu \right) \exp \left[
2\int_{r_0}^r\left( \frac 1r-\frac 1{r-2\mu }\right) dr\right] \quad .
\end{equation}
Here $r_0$ is a fiducial radius that can be chosen as appropriate to a
physical situation. Thus if $r_0\rightarrow \infty $ we expect to recover
the Schwarzschild case. The latter is indeed recovered for $\lambda =0$,
when $\mu =M=$ constant and (13) reads

\begin{equation}
B=A\exp \left[ 2\int_\infty ^r\frac{A-1}rdr\right] =1-\frac{2M}r\;\;.
\end{equation}
That is, $\lambda =0$ specifies the Schwarzschild limit of a class of
solutions determined by (16) and one or the other of (11) and (12).

\section{CLOSE-TO-SCHWARZSCHILD APPROXIMATION}

We can study close-to-Schwarzschild approximation by expanding the mass
function $\mu (r)$ of (15) for $\left| \lambda \right| $ small (i.e. $\left|
\lambda \right| ^{-1}\gg M)$. Thus we write

\begin{equation}
\mu (r)=\sum_{n=0}^\infty \mu _n(r)
\end{equation}
with $\mu _0=M$. Here $\mu _0$ is the zero-order approximation of $\mu $, $%
(\mu _0+\mu _1)$ is the first-order approximation, and so on. To obtain $\mu
_1$ we substitute $\mu _0$ for $\mu $ in the RHS of (16) to obtain $\mu
_1^{\prime }$. Integrating $\mu _1^{\prime }$ gives $\mu _1$. Then we
substitute $(\mu _0+\mu _1)$ for $\mu $ in the RHS of (16) to obtain $\mu
_2^{\prime }$. In this way we obtain the following recursion formulae for
evaluating $\mu _n(r)$: 
\begin{eqnarray}
\mu _0^{\prime } &=&0  \nonumber \\
\mu _1^{\prime } &=&-\mu _0^{\prime }+\frac 32\lambda ^2r\left( r-2M\right)
\exp \left[ 2\int_{r_0}^r\left( \frac 1r-\frac 1{r-2M}\right) dr\right] 
\nonumber \\
&&..............  \nonumber \\
\mu _n^{\prime } &=&-\sum_{m=0}^{n-1}\mu _m^{\prime }+\frac 32\lambda
^2r\left( r-2\sum_{m=0}^{n-1}\mu _m\right) .  \nonumber \\
&&\exp \left\{ 2\int_{r_0}^r\left[ \frac 1r-\left( r-2\sum_{m=0}^{n-1}\mu
_m\right) ^{-1}\right] dr\right\} \quad .
\end{eqnarray}
From this we find

\begin{equation}
\mu _1^{\prime }=\frac 32\lambda ^2\left( 1-\frac{2M}{r_0}\right) ^2\,\frac{%
r^3}{r-2M}\;,
\end{equation}
and so

\begin{equation}
\mu _1=\frac 32\lambda ^2\;\left( 1-\frac{2M}{r_0}\right) ^2\int_{r_0}^r%
\frac{r^3dr}{r-2M}\;.
\end{equation}
In the region $r\gg 2M$ and $r_0\gg 2M$, we find

\begin{equation}
\mu _1\thickapprox \frac 12\lambda ^2r^3
\end{equation}
where we have absorbed a constant term in $\mu _0$ without loss of
generality, so the first-order approximation of $A(r)$ in (15) is

\begin{equation}
A\thickapprox \left( 1-\frac{2M}r-\lambda ^2r^2\right) ^{-1}.
\end{equation}
Substituting this equation into (13), keeping only terms up to first order
in $\lambda ^2$ as well as in $M$, and neglecting a constant factor in $B$,
we find

\begin{equation}
B\thickapprox 1-\frac{2M}r+2\lambda ^2r^2\;.
\end{equation}
From this coefficient and that of (23) we obtain the first order
close-to-Schwarzschild approximation of the 5D solution as 
\begin{equation}
dS^2\approx e^{2\lambda t}\left[ \left( 1-\frac{2M}r+2\lambda ^2r^2\right)
dt^2-\left( 1-\frac{2M}r-\lambda ^2r^2\right) ^{-1}dr^2-r^2d\Omega ^2\right]
-e^{-4\lambda t}dy^2.
\end{equation}
This metric is time-dependent. When $\lambda \rightarrow 0$, it tends to the
5D Schwarzschild solution. So we call (25) the close-to-Schwarzschild
approximation for small $\left| \lambda \right| ,$ implying that the
time-variation of the field should be very slow. We also find that the
solution (25) belongs to the Type D class of the general time-dependent 5D
metrics.$^{22}$ As regards the 4D part inside the square bracket in (25), it
is interesting to note that it does \underline{not} define the 4D
Schwarzschild-de Sitter solution. In the latter, $A$ and $B$ both contain a
term $\Lambda r^2/3$ where $\Lambda $ is the cosmological constant. By
contrast, whereas $\lambda ^2$ has the same physical dimensions as $\Lambda $
(namely $time^{-2}$ or $length^{-2}$), $A$ and $B$ in (23) and (24) contain
terms with different signs and different sizes. This situation is analogous
to another in Kaluza-Klein theory, where the exact solution of the 5D field
equations for a charged point mass does not exactly reproduce the 4D
Reissner-Nordstrom solution.$^{23}$ Since the 5D equations are richer than
the 4D ones, such situations may be expected; but even so, (25) defines a
new solution.

\section{TWO EXACT 4D SOLUTIONS}

There are, of course, many other solutions than those which are close to
Schwarzschild. In general, we have a class of solutions of $R_{AB}=0$ with a
metric

\begin{equation}
dS^2=e^{2\lambda t}\left[ Bdt^2-Adr^2-r^2d\Omega ^2\right] -e^{-4\lambda
t}dy^2
\end{equation}
whose spacetime potentials $A,B$ are determined by (11),(12). Alternatively,
they are determined by (16) and one or the other of (11), (12). We should
recall, however, that while (8), (9), (10) are 4D relations and can be used
as such, the class of metrics (26) is 5D in nature. This means that we can
use Campbell's theorem$^1$ and the induced-matter formalism$^7$ to generate
4D solutions of Einstein's equations with their appropriate matter. We now
proceed to show how the 5D metric (26) produces two exact 4D solutions.

\subsection{SOLUTION WITH A STIFF FLUID}

The first is obtained by splitting off the part inside square brackets in
(26). The 4D metric is then

\begin{equation}
ds^2=g_{\alpha \beta }dx^\alpha dx^\beta =Bdt^2-Adr^2-r^2d\Omega ^2\quad ,
\end{equation}
and is static. We therefore expect that the source necessary to balance
Einstein's equations will also be static. As mentioned above, the
Kaluza-Klein equations $R_{AB}=0\;(A,B=0,123,4)$ contain as a subset the
Einstein equations $R_{\alpha \beta }-Rg_{\alpha \beta }/2=T_{\alpha \beta }$
($\alpha ,\beta =0,123$). Here $T_{\alpha \beta }$ is the 4D energy-momentum
tensor, which can always be constructed from the 5D geometry$^{1,2}$ and
whose form is now well known.$^{7-9}$ For the present case, $R_{\alpha \beta
}$ is given by (3), so the induced energy-momentum tensor is

\begin{equation}
T_{\alpha \beta }=R_{\alpha \beta }-\frac 12g_{\alpha \beta }R=6\Phi
^{-2}\left( \Phi _\alpha \Phi _\beta -\frac 12g_{\alpha \beta }\Phi ^\mu
\Phi _\mu \right) .
\end{equation}
Using (7) and (27), we find that the non-vanishing components of this in
mixed form are

\begin{equation}
T_0^0=-T_1^1=-T_2^2=-T_3^3=3\lambda ^2B^{-1}\quad .
\end{equation}
This as expected represents a static source, which we can model as a perfect
fluid with

\begin{equation}
T_{\alpha \beta }=\left( \rho +p\right) u_\alpha u_\beta -pg_{\alpha \beta
}\quad .
\end{equation}
Here $\rho $ is the density, $p$ is the pressure and the 4-velocity is $%
u^\alpha =(u^0,0,0,0)$. Combining (29) and (30), we see that the source has

\begin{equation}
\rho =p=3\lambda ^2B^{-1}\quad .
\end{equation}
This is the stiff equation of state in which the speed of sound waves
approaches the speed of light, and has been applied in previous studies$%
^{14,16,21}$ to astrophysical situations such as collapsed stars and
proto-galactic fluctuations.

\subsection{TWO-FLUID SOLUTION WITH RADIATION AND HEAT FLOW}

The second solution we look at is obtained by splitting off the whole of the
first part of (26). \ The 4D metric is then

\begin{equation}
d\tilde{s}^2=\tilde{g}_{\alpha \beta }dx^\alpha dx^\beta =e^{2\lambda
t}\left[ Bdt^2-Adr^2-r^2d\Omega ^2\right] ,
\end{equation}
and is time-dependent. We therefore expect that the source necessary to
balance Einstein's equations will also be time-dependent. It should be noted
that this property cannot in general be removed by a coordinate
transformation based on Birkhoff'sss theorem, because the 4D metric (32)
will have a source constructed from the 5D geometry that will in general not
be vacuum; and because the 4D metric (32) is part of a 5D metric (26) and it
is known that Birkhoff's theorem in its standard form breaks down in the
transition from 4D to 5D,$^{24,25}$ as evidenced by the existence of both
static and time-dependent soliton solutions.$^{24,26}$ To investigate the
time-dependence of (32), let us make the coordinate transformation

\begin{equation}
e^{\lambda t}=1-\lambda \tilde{t}\quad .
\end{equation}
This brings (32) into the form

\begin{equation}
d\tilde{s}^2=\tilde{g}_{\alpha \beta }d\tilde{x}^\alpha d\tilde{x}^\beta =Bd%
\tilde{t}^2-\left( 1-\lambda \tilde{t}\right) ^2\left( Adr^2+r^2d\Omega
^2\right) \quad .
\end{equation}
We can use this with $\phi \equiv \Phi ^{-2}=e^{-2\lambda t}=(1-\lambda 
\tilde{t})^{-2}$ to evaluate the induced matter properties using the
standard technique.$^{7-9}$ We note that

\begin{eqnarray}
\tilde{R}_{\alpha \beta } &=&\phi ^{-1}\widetilde{\phi _{\alpha ;\beta }} \\
\tilde{R} &=&\phi ^{-1}\widetilde{\phi _{;\alpha }^\alpha }=0\;\;,
\end{eqnarray}
which can be used to form the induced energy-momentum tensor:

\begin{equation}
\tilde{T}_{\alpha \beta }=\tilde{R}_{\alpha \beta }-\frac 12\tilde{g}%
_{\alpha \beta }\tilde{R}=\phi ^{-1}\widetilde{\phi _{\alpha ;\beta }}\quad .
\end{equation}
The non-vanishing components of this in mixed form are

\begin{eqnarray}
\tilde{T}_0^0 &=&6\lambda ^2B^{-1}\left( 1-\lambda \tilde{t}\right) ^{-2} \\
\tilde{T}_1^1 &=&\tilde{T}_2^2=\tilde{T}_3^3=-2\lambda ^2B^{-1}\left(
1-\lambda \tilde{t}\right) ^{-2} \\
\tilde{T}_0^1 &=&g_{00}g^{11}\tilde{T}_1^0=\lambda A^{-1}B^{-1}B^{\prime
}\left( 1-\lambda \tilde{t}\right) ^{-3}\quad .
\end{eqnarray}
We see from (38) that the density is inhomogeneous via $B=B(r)$ and is
time-dependent, as is the pressure by (39) though the latter is isotropic.
We also see from (40) that there is an off-diagonal component. The latter
can be accommodated by introducing a two-fluid model

\begin{equation}
\tilde{T}_{\alpha \beta }=\tilde{T}_{\alpha \beta }^{(1)}+\tilde{T}_{\alpha
\beta }^{(2)}\quad ,
\end{equation}
which as in other work$^{12,16}$ we take to be the sum of a perfect fluid
and a radial heat flow:

\begin{eqnarray}
\tilde{T}_{\alpha \beta }^{(1)} &=&\left( \tilde{\rho}+\tilde{p}\right)
u_\alpha u_\beta -\tilde{p}\tilde{g}_{\alpha \beta } \\
\tilde{T}_{\alpha \beta }^{(2)} &=&q_\alpha u_\beta +u_\alpha q_\beta \quad .
\end{eqnarray}
The heat-flux vector and the 4-velocity must obey the orthogonality condition

\begin{equation}
q_\alpha u^\alpha =0\quad .
\end{equation}
This we satisfy by taking

\begin{eqnarray}
u^\alpha &=&\left( u^0,0,0,0\right) \;,\;\;\;\;\;\;\;u^0=B^{-1/2} \\
q^\alpha &=&\left( 0,q^1,0,0\right) \quad \quad ,
\end{eqnarray}
where in (45) we have used the facts that $r$ is a comoving coordinate and
that the 4-velocities are normalized via $g_{\alpha \beta }u^\alpha u^\beta
=Bu^0u^0=1$. Substituting (45), (46) into (42), (43) and these into (41)
gives the non-vanishing components of the last in mixed form:

\begin{eqnarray}
\tilde{T}_0^0 &=&\tilde{\rho} \\
\tilde{T}_1^1 &=&\tilde{T}_2^2=\tilde{T}_3^3=-\tilde{p} \\
\tilde{T}_0^1 &=&B^{1/2}q^1\quad .
\end{eqnarray}
Comparing these with (38), (39), (40) gives us the density, pressure and
heat flow in explicit form:

\begin{eqnarray}
\tilde{\rho} &=&3\tilde{p}=6\lambda ^2B^{-1}\left( 1-\lambda \tilde{t}%
\right) ^{-2} \\
q^1 &=&\lambda A^{-1}B^{-3/2}B^{\prime }\left( 1-\lambda \tilde{t}\right)
^{-3}\quad .
\end{eqnarray}
The equation of state is that of radiation or ultra-relativistic matter, and
has been applied in previous studies$^{16,27}$ to astrophysical situations
such as fermion soliton stars and the early universe.

\section{DISCUSSION AND CONCLUSION}

We have taken a 5D metric (1) which contains a 4D sub-metric (6) whose 3D
part is spherically symmetric. An expansion in the metric coefficient allows
us to recover the Schwarzschild case of general relativity in (17) as the
zeroth approximation and a close-to-Schwarzschild case in (25) as the first
order approximation to a class of spherically-symmetric solutions. Any
solution of the 5D Kaluza-Klein equations in apparent vacuum can be written
as a solution of the 4D Einstein equations with matter. Two solutions have
then been shown, with matter properties corresponding to those of a stiff
fluid (31), and radiation or ultra relativistic particles with heat flow
(50,51). These solutions can be applied to astrophysics, but are merely
illustrative examples.

In the wider scheme, it is clear that (local) embedding theorems are
powerful tools, whether applied to $N<4$ (possibly quantizeable) gravity,$^2$
4D Einstein theory,$^{12}$ 5D Kaluza-Klein theory,$^{28}$ or 10D
superstrings, 11D supergravity and M-theory.$^{3-6}$ Campbell's theorem
ensures that any solution in\ ND can be embedded in a Ricci-flat solution in 
$(N+1)D$ .$^1$ The Schwarzschild solution in 4D can of course be embedded in
a flat space of $N\geq 6$ .$^{29}$ And any solution in 4D can be embedded in
a flat space of $N\geq $ $10$ .$^{30}$ The implications of embedding
theorems are diverse. If the aim is to find new solutions of general
relativity, the higher-dimensional field equations are often surprisingly
tractable, and the method of reduction to 4D is straightforward.$^2$ If the
aim is to give meaning to higher-dimensional theories, the same method of
reduction will inform about physicality in the 4D world. We should recall
that field equations like those of Einstein or Kaluza-Klein allow the
dimensionality to be freely chosen, which should be done partly with a view
to what physics it is desired to describe and partly with a view to what
technical constraints are involved. In the latter regard, it is well known
that $N<4$ theories run into problems of formulation which are connected
with the degenerate nature of lower-dimensional Riemannian spaces. For $N>4$
theories, problems arise with the physics if the spaces are subject to
arbitrary technical constraints.$^{7,31,32}$ Our opinion, therefore, is that
future work should be focussed on higher-dimensional, fully covariant theory.

\section*{ACKNOWLEDGMENTS}

We thank the Referee for valuable suggestions, and J. Ponce de Leon and W.N.
Sajko for comments. This work was supported by NSF of P.R. China and NSERC of
Canada.


\begin{thebibliography}{99}
\bibitem{}  J.E. Campbell, A Course of Differential Geometry, Clarendon,
Oxford (1926).

\bibitem{}  S. Rippl, C. Romero, R. Tavakol, Class. Quant. Grav. {\bf 12},
2411 (1995); \ C. Romero, R. Tavakol, R. Zalaletdinov, Gen. Rel. Grav. {\bf %
28}, 365 (1996); \ J.E. Lidsey, C. Romero, R. Tavakol, S. Rippl, Class.
Quant. Grav. {\bf 14}, 865 (1997).

\bibitem{}  E. Witten, Nucl. Phys. B {\bf 186}, 412 (1981);\ P. Horava, E.
Witten, Nucl. Phys. B {\bf 460}, 506 (1996); \ {\bf 475}, 94 (1996).

\bibitem{}  P. West, Introduction to Supersymmetry and Supergravity, World
Scientific, Singapore (1986).

\bibitem{}  M.B. Green, J.H. Schwarz, E. Witten, Superstring Theory,
Cambridge Un. Press, Cambridge (1987).

\bibitem{}  M.J. Duff, Int. J. Mod. Phys. A {\bf 11}, 5623 (1996).

\bibitem{}  P.S. Wesson, Space, Time, Matter, World Scientific, Singapore
(1999).

\bibitem{}  P.S. Wesson, Astrophys. J. {\bf 394}, 19 (1992).

\bibitem{}  A. Billyard, P.S. Wesson, Phys. Rev. D {\bf 53}, 731 (1996).

\bibitem{}  D. Kalligas, P.S. Wesson, C.W.F. Everitt, Astrophys. J. {\bf 439}%
, 548 (1995); \ H. Liu, J.M. Overduin, Astrophys. J. {\bf 538}, 386 (2000);
\ J.M. Overduin, Phys. Rev. D {\bf 62}, 102001 (2000).

\bibitem{}  B. Mashhoon, H. Liu, P.S. Wesson, Phys. Lett. B {\bf 331}, 305
(1994); \ H. Liu, B. Mashhoon, Ann. Phys. (Leipzig) {\bf 4}, 565 (1995); \
P.S. Wesson, B. Mashhoon, H. Liu, Mod. Phys. Lett. A {\bf 12}, 2309 (1997);
\ H. Liu, B. Mashhoon, Phys. Lett. A {\bf 272}, 26 (2000).

\bibitem{}  D. Kramer, H. Stephani, M. MacCallum, E. Herlt, Exact Solutions
of Einstein's Field Equations, Cambridge Un.\ Press, Cambridge (1980).

\bibitem{}  P.C. Vaidya, Nature (London) {\bf 171}, 260 (1953).

\bibitem{}  L. Herrera, J. Jimenez, L. Leal, J. Ponce de Leon, M. Esculpi,
V. Galina, J. Math. Phys. {\bf 25}, 3274 (1984).

\bibitem{}  L. Herrera, J. Ponce de Leon, J. Math. Phys. {\bf 26}, 2302
(1985); \ {\bf 26}, 2847 (1985).

\bibitem{}  M. Esculpi, L. Herrera, J. Math. Phys. {\bf 45}, 3341 (1992).

\bibitem{}  E. Glass, J.P. Krisch, Phys. Rev. D {\bf 57}, R5945 (1998).

\bibitem{}  G. Abolghasem, A.A. Coley, D.J. McManus, J. Math. Phys. {\bf 37}
, 361 (1996); \ H. Liu, P.S. Wesson, Gen. Rel. Grav. {\bf 30}, 509 (1998); \
P.S. Wesson, H. Liu, Phys. Lett. B {\bf 432}, 266 (1998).

\bibitem{}  T. Appelquist, A. Chodos, Phys. Rev. D {\bf 28}, 772 (1983); \
H. Liu, P.S. Wesson, Int. J. Mod. Phys. D {\bf 7}, 737 (1998).

\bibitem{}  C.W. Misner, D.M. Sharp, Phys. Rev. B {\bf 136}, 571 (1964); \
M.P. Podurets, Sov. Astron. (A.J.) {\bf 8}, 19 (1964); \ P.S. Wesson, Mon.
Not. Roy. Astr. Soc. {\bf 197}, 157 (1981).

\bibitem{}  P.S. Wesson, J. Math. Phys. {\bf 19}, 2283 (1978).

\bibitem{}  H. Liu, P.S. Wesson, J. Ponce de Leon, J. Math. Phys. {\bf 34},
4070 (1993).

\bibitem{}  H. Liu, P.S. Wesson, Phys. Lett. B {\bf 381}, 420 (1996); \
Class. Quant. Grav. {\bf 14}, 1651 (1997).

\bibitem{}  D.J. Gross, M.J. Perry, Nucl. Phys. B {\bf 226}, 29 (1983).

\bibitem{}  Y.M. Cho, D.H. Park, Gen. Rel. Grav. {\bf 23}, 741 (1991); \
P.S. Wesson, {\bf in} STEP: Testing the Equivalence Principle in Space,
European Space Agency, WPP-115, 566 (1996).

\bibitem{}  R.D. Sorkin, Phys. Rev. Lett. {\bf 51}, 87 (1983); \ A.
Davidson, D. Owen, Phys. Lett. B {\bf 155}, 247 (1985); \ P.S. Wesson, H.
Liu, P. Lim, Phys. Lett. B {\bf 298}, 69 (1993).

\bibitem{}  T.D. Lee, Phys. Rev. D {\bf 35}, 3637 (1987); \ R. Frieberg,
T.D. Lee, Y. Pang, Phys. Rev. D {\bf 35}, 3640 (1987); \ T.D. Lee, Y. Pang,
Phys. Rev. D {\bf 35}, 3678 (1987); \ R. Frieberg, T.D. Lee, Y. Pang, Phys.
Rev. D {\bf 35}, 3658 (1987); \ P.S. Wesson, H. Liu, Astrophys. J. {\bf 440}%
, 1 (1995).

\bibitem{}  T. Kaluza, Sitz, Preuss. Akad. Wiss. {\bf 33}, 966 (1921); \ O.
Klein, Z. Phys. {\bf 37}, 895 (1926); \ J.M. Overduin, P.S. Wesson, Phys.
Rep. {\bf 283}, 303 (1997).

\bibitem{}  F.R. Tangherlini, Nuovo Cimento {\bf 27} (Ser. 10), 636 (1963).

\bibitem{}  L.P. Eisenhart, Riemannian Geometry, Princeton Un. Press,
Princeton (1949).

\bibitem{}  N. Arkani-Hamed, S. Dimopoulos, G. Dvali, Phys. Lett. B {\bf 429}%
, 263 (1998); \ Phys. Rev. D {\bf 59}, 086004 (1999); \ I. Antoniadis, N.
Arkani-Hamed, S. Dimopoulos, G. Dvali, Phys. Lett. B {\bf 436}, 257 (1998).

\bibitem{}  L. Randall, R. Sundrum, Phys. Rev. Lett. {\bf 83}, 3370 (1999);
\ {\bf 83}, 4690 (1999).
\end{thebibliography}
\end{document}